\documentclass[prl,showpacs,superscriptaddress]{revtex4}
\usepackage{amsmath,mathrsfs}
\usepackage{epsfig}
\begin{document}

\title{Spintronic transport and Kondo effect in quantum dots}
\author{David S\'anchez}%
\author{Rosa L\'opez}%
\affiliation{D\'epartement de Physique Th\'eorique,
Universit\'e de Gen\`eve, CH-1211 Gen\`eve 4, Switzerland}%
\author{Mahn-Soo Choi}%
\affiliation{Department of Physics, Korea University, Seoul 136-701,
Korea}%

\date{\today}

\begin{abstract}
We investigate the spin-dependent transport properties of quantum-dot
based structures where Kondo correlations dominate the electronic
dynamics. The coupling to ferromagnetic leads with parallel
magnetizations is known to give rise to nontrivial effects in the local
density of states of a single quantum dot. We show that this influence
strongly depends on whether charge fluctuations are present or absent in
the dot.  This result is confirmed with numerical renormalization group
calculations and perturbation theory in the on-site interaction.  In the
Fermi-liquid fixed point, we determine the correlations of the electric
current at zero temperature (shot noise) and demonstrate that the Fano
factor is suppressed below the Poissonian limit for the symmetric point
of the Anderson Hamiltonian even for nonzero lead magnetizations.
We discuss possible avenues of future research in this field: coupling
to the low energy excitations of the ferromagnets (magnons), extension
to double quantum dot systems with interdot antiferromagnetic
interaction and effect of spin-polarized currents on higher symmetry
Kondo states such as SU(4).
\end{abstract}

\pacs{Kondo effect, quantum dots, spin-polarized transport,
  spin-dependent tunneling}

\maketitle

\section{Introduction}

The study of spin-polarized transport across interfaces is a subject of
long history~\cite{ted71,jul75}.  The recent advent of
semiconductor-based electronic devices at the nanoscale has revived the
interest in transferring, controlling and detecting spin currents. This
research area has been termed {\em spintronics} due to the exciting
possibility of future, successful spin-based electronic
technology~\cite{book}.  Nevertheless, spintronics is interesting as
well for fundamental physics, both experimentally and theoretically, as
its basic constituent---the spin---is of quantum nature only.

The most simple building block of spintronic transport systems is
probably the magnetic tunnel junction. It comprises two ferromagnetic
electrodes sandwiching a paramagnetic layer. Vertical transport, where
current flows across the interfaces, is characterized by the tunneling
magnetoresistance (TMR), which measures the relative change in the
junction resistance when the contacts' magnetizations is changed from
parallel to antiparallel alignment~\cite{pri98}. The TMR is also
important in investigating the properties of spintronic resonant
tunneling diodes. Remarkably, such devices have been recently built in
all-semiconductor heterostructures, taking advantage of diluted magnetic
semiconductors made of III-V~\cite{ohn98,mak98,hay00} (hole-like
transport) and II-VI compounds~\cite{slo03} (electron-like transport).

When the size of the paramagnetic island in a magnetic tunnel junction
becomes comparable to the carrier Fermi wavelength, the system behaves
effectively as zero-dimensional. Then, quantum effects arise from the
quasi-localized nature of electrons and from the phase-coherent
transport. The ultimate miniaturization limit is just a single resonant
level coupled to a Fermi sea of itinerant electrons, which may be
regarded as an artificial realization of the quantum {\em impurity
  problem}~\cite{mah83}.  Extensive studies of the impurity problem have
been performed in semiconductor {\em quantum dots}~\cite{kou97}, where
the island (an electron droplet) is formed by means of a constriction in
a two-dimensional electron gas. Both the discrete energy levels and the
tunneling couplings may be tuned almost at will.  Very recently, a few
experimental works have begun to deal with spin polarized
leads~\cite{haw99,kel01}.

In this paper, we consider spintronic transport through quantum dots
with strong correlations. It is well known that the electron-electron
interaction interaction plays a dominant role in the low temperature
transport through quantum dots~\cite{kou97}.  In the Coulomb blockade
regime, the electron dynamics can be described in terms of
single-electron tunneling plus mean-field charging effects.  Between two
Coulomb blockade peaks, transport is blockaded and the electron number
in the dot does not fluctuate.  When the electron number is odd, the
topmost resonant level is singly occupied. However, when temperature
approaches the energy scale $T\sim T_K$, the spin of the localized
electron becomes screened by an antiferromagnetic interaction with the
conduction band electrons. The resulting strong correlations arise from
the interplay of higher order tunneling processes and the on-site
Coulomb interaction. As a consequence, the many-body state in the $T=0$
limit is a singlet formed between the quantum impurity and the continuum
electrons.  The impact in the transport properties of the system is
strong; e.g., the temperature-dependent conductance results in a
universal function of $T/T_K$, achieving the quantum limit ($e^2/h$ per
spin) at $T=0$. This is the celebrated Kondo effect~\cite{hew93} in
quantum dots~\cite{gla88,ng88,kou01}, which was observed several years
ago~\cite{exp}.

Now, an important condition for the Kondo effect to take place is the
degeneracy (between spins up and down) of the ground state of the dot.
Such spin degeneracy may be broken with an external magnetic field
(Zeeman splitting) and is well understood~\cite{mei93,cos00}. But will
the Kondo effect be preserved when the spin transfer across the tunnel
barrier is spin-dependent?  How will the conventional picture of the
Kondo resonance in quantum dots be affected within a spin-polarized
medium?  To answer these questions various theoretical groups have
lately
contributed~\cite{ser02,zha02,bul03,lop03,mar03,don03,mar04,choi04,%
  lu02,ma02,ma03,tan04}.
Although these works differ in some predictions due to the range of
applicability of the distinct approaches used therein and their
limitations, the analysis of the problem with numerical renormalization
group~\cite{mar04,choi04} have led to the conclusion that the Kondo
state is robust enough (though with a lower $T_K$) against nonzero spin
polarizations in the leads when particle-hole symmetry is not broken and
real charge fluctuations are completely suppressed~\cite{choi04}.  Note
that particle-hole asymmetry may be induced in the dot with nearby
electric gates, shifting the resonant level away from the symmetric
point (see below)~\cite{mar04b}.  We predict that this would give rise
to a sharp decrease of the linear-response conductance. Since the
phenomenon is absent when the magnetizations of the electrodes point
along opposite directions (antiparallel alignment), we
propose~\cite{choi04} the TMR as a possible experimental signature of
this {\em spintronic Kondo effect}.

Another important element of many theories of spintronic transport is
the description of intrinsic ``spin relaxation'' mechanisms that allow
for nonequilibrium spin populations to relax.  Long spin coherence times
$\tau_{\rm sf}$ have been reported in semiconductor quantum
wells~\cite{kik97} and dots~\cite{han03}. The effect of spin relaxation
is known to reduce the TMR for a Coulomb-blockaded quantum
dot~\cite{rud01} and it leads to a suppression of the Fano factor (shot
noise) in the antiparallel configuration~\cite{sou02}.  At lower
temperatures ($T<T_K$), spin decoherence causes the destruction of the
Kondo effect due the failure of the formation of the many-body singlet
state.
One could also think about more coherent ``spin-flip'' process, e.g.,
arising from the potential spin-orbit coupling which causes the rotation
of electron spin in the dot (this effect of spin-orbit coupling for the
localized electron should be distinguished from that of the spin-orbit
coupling of conduction electrons, which due to time-reversal symmetry
has no influence on the Kondo effect~\cite{Meir94a}).
More specifically, when the amplitude of spin flip scattering rate is
larger than the Kondo temperature, $h/2\tau_{\rm sf}\gtrsim T_K$, the
density of states (DOS) at the impurity site is expected to develop a
splitting and thus a decrease of the linear conductance.
This prediction has been confirmed with equation-of-motion
technique~\cite{zha02}, slave-boson mean-field theory~\cite{lop03,ma03}
and numerical renormalization group~\cite{choi04}.
Therefore, spin-flip processes, coherent or incoherent, tend to suppress
the Kondo effect.

\section{Hamiltonian and Theoretical Approaches}
\begin{figure}\centering
\includegraphics*[width=85mm]{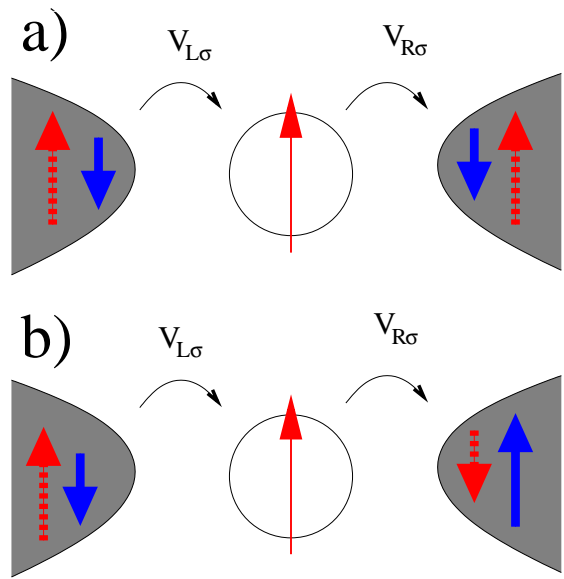}
\caption{(Color online). Schematic picture of the quantum dot attached
  to ferromagnetic leads. The dot is a single resonant 
  level which may be shifted through a capacitative coupling to a gate.
  In (a) we show the parallel configuration. Dashed arrows denote the
  majority spins and solid arrows are for minority ones. Case (b)
  corresponds to the antiparallel alignment.}
\label{scheme}
\end{figure}

We model the quantum dot as a single discrete level with energy
$\varepsilon_{d,\sigma}$ containing an unpaired spin-$1/2$ electron with
$\sigma=\{\uparrow,\downarrow\}$ and charging energy $U$.  Therefore,
the dot is an electronic impurity tunnelling coupled to continuum
electrons with a model Hamiltonian given by the Anderson Hamiltonian:
\begin{equation}\label{eq_h}
{\cal H}={\cal H}_{\rm leads}+{\cal H}_{\rm dot}+{\cal H}_{\rm coupling}\,,
\end{equation}
where (see Fig.~\ref{scheme})
\begin{align}
{\cal H}_{\rm leads}=\sum_{k\alpha \sigma}\varepsilon_{k\alpha \sigma}
c_{k\alpha \sigma}^\dagger c_{k\alpha \sigma} \,,\\
{\cal H}_{\rm dot}=\sum_{\sigma}\varepsilon_{d,\sigma}\hat{n}_\sigma +U
\hat{n}_\uparrow \hat{n}_\downarrow +
(R d_\uparrow^\dagger d_\downarrow +{\rm H.c.}) \label{eq_hr}\,,\\
{\cal H}_{\rm coupling}=\sum_{k\alpha \sigma} (V_{k\alpha \sigma}
c_{k\alpha \sigma}^\dagger d_{\sigma}+ {\rm H.c.}) \,,
\end{align}
are written in terms of the creation and annihilation operators in the
dot $d_\sigma^\dagger,d_\sigma$ (the occupation number is defined as
$\hat{n}_\sigma=d_\sigma^\dagger d_\sigma$) and in the leads $c_{k\alpha
  \sigma}^\dagger,c_{k\alpha\sigma}$, with $k$ the wavevector and
$\alpha$ labeling left ($\alpha=L$) and right ($\alpha=R$) reservoirs.
Tunneling of electrons from the dot to the leads is described by the
hopping parameter $V_{\alpha k\sigma}$.
In ${\cal H}_{\rm dot}$ included is an internal spin-flip process with
rate $\tau_{\rm sf}^{-1}\sim 2R/\hbar$~\cite{rud01,sou02}.
Notice that in this framework the spin-flip process is purely coherent,
and precisely speaking it does not account for incoherent spin
relaxation processes.  It may originate either from the transverse
component of an applied magnetic field or from a tunable spin-orbit
coupling of the Rashba-Dresselhaus type \emph{in the dot}~\cite{book}
(see the Introduction and compare with Ref.~\cite{Meir94a}).
Since the spin-flip processes, coherent or incoherent, have similar
influence on Kondo effect, we leave the term in $R$ in
$\mathcal{H}_\mathrm{dot}$ phenomenological.
What is important here is that $R$ lifts the degeneracy of the discrete
level and that it cannot be eliminated with a unitary transformation
since the lead magnetizations already mark a privileged spin direction.
For $R=0$ and $p\neq 0$, the SU(2) symmetry is broken and the spin
symmetry of the problem is U(1) whereas in the presence of both spin
flip scattering and ferromagnetic electrodes, the U(1) spin symmetry is
explicitly broken.

Due to coupling to the leads, the electron in the dot becomes
quasilocalized with a escaping rate related to the hybridization
broadening, $\Gamma_{\alpha\sigma}(\omega)= \pi \sum_k |V_{k\alpha
  \sigma}|^2\delta(\omega-\varepsilon_{k\alpha \sigma})$.  This is the
imaginary part of the hopping self-energy, which is spin-dependent
because tunneling is spin-dependent. This can be achieved by coupling
the dot to ferromagnetic leads.  We take constant tunneling coefficients
$V_\alpha$ and equal tunnel barriers (symmetric couplings: $V_L=V_R$).
In the wide-band limit, the energy dependence of
$\Gamma_{\alpha\sigma}(\omega)$ is unimportant (which is a good
approximation for low voltages). Moreover, we assume that the degree of
spin polarization on lead $\alpha$ is given by
\begin{equation}\label{eq_p}
p_\alpha=
\frac{\Gamma_{\alpha\uparrow}-\Gamma_{\alpha\downarrow}}{
\Gamma_{\alpha\uparrow}+\Gamma_{\alpha\downarrow}}
\,,
\end{equation}
Notice that Eq.~(\ref{eq_p}) is already a gross simplification as it
might well be that $p_\alpha$ has little to do with the real
magnetization of the reservoir. In fact, various definitions for $p$ are
possible depending on the experiment~\cite{maz99}.  In addition, we
neglect proximity effects such as stray fields coming from the
ferromagnets
and consider that the bandwidth $D$ is spin independent.
[We prefer not to delve into details since already the simple form of
Eq.~(\ref{eq_p}) gives rise to nontrivial effects which can be directly
measured].

We consider collinear magnetizations, both in parallel (P) and
antiparallel (AP) configurations.  With the approximations discussed
above, we have for the P case ($p_L=p_R\equiv{p}$) $\Gamma_{L\uparrow} =
\Gamma_{R\uparrow} = (1+p)\Gamma_0/2$ and $\Gamma_{L\downarrow} =
\Gamma_{R\downarrow} = (1-p)\Gamma_0/2$, where
$\Gamma_0\equiv\Gamma_{\alpha\uparrow}+\Gamma_{\alpha\downarrow}$,
whereas the AP case ($p_L=-p_R\equiv{p}$) yields $\Gamma_{L\uparrow} =
\Gamma_{R\downarrow} = (1+p)\Gamma_0/2$ and $\Gamma_{L\downarrow} =
\Gamma_{R\uparrow} = (1-p)\Gamma_0/2$.

Let us first provide an intuitive picture of the influence of
spin-polarized transport in the Kondo resonance at $R=0$.  We take
$E_F=0$.  For P alignment in the fully polarized case ($p_L=p_R=1$), the
singlet state cannot form due to the lack of spin down electrons. Hence,
we expect a decrease of the Kondo temperature ($T_K$ is roughly the
binding energy of the singlet state) with increasing $p$.  In the AP
configuration ($p_L=-p_R=1$), however, the Kondo effect survives since
an spin up (down) localized electron may be screened by the right (left)
electrode. Of course, the conductance would be zero unless a vanishingly
small $R$ is allowed to come into play. Now, in the P case there may
arise an exchange field~\cite{mar03} acting on the dot as an effective
Zeeman splitting~\cite{choi03}. Is the Kondo effect robust against this
exchange field?  The answer is yes!~\cite{choi04}. When the gate voltage
is tuned in such a way that $\varepsilon_d=-U/2$ (the symmetric Anderson
model), charge fluctuations become suppressed. Only when particle-hole
symmetry is broken ($\varepsilon_d\neq -U/2$) do we find a splitting in
the Kondo peak of the local DOS.

We briefly review now the different theoretical methods employed to
solve Eq.~(\ref{eq_h}). The equation-of-motion technique~\cite{ng96} is
useful to study nonequilibrium situations (for finite bias) at
relatively ``high'' temperatures ($T\gtrsim T_K$).  Although it
reproduces qualitatively the DOS peaks, it fails to describe properly
the strong coupling regime, where Kondo physics completely quenches the
impurity spin. When applied to our problem, it predicts the exchange
field induced splitting but not its disappearance at
$\varepsilon_d=-U/2$.  On the other hand, slave-boson mean-field
theory~\cite{col84} correctly accounts for the Fermi-liquid fixed point
of the Kondo problem at $T=0$. As a result, it is only valid when the
particle-hole symmetry is not broken (no splitting).  The noncrossing
approximation~\cite{bic87} is another slave-boson based approach and
offers a consistent picture of the Kondo effect at $T\sim T_K$.
However, it does not take into account vertex corrections and produces
spurious peaks at $E_F$ in the presence of a magnetic field. Finally, a
numerical renormalization group calculation~\cite{kri80} encompasses the
whole regime but remains valid only at equilibrium.

In the following, we report results using the interpolative $U$-finite
perturbation theory~\cite{kaj} since it gives a good description of the
dynamical properties of Eq.~(\ref{eq_h}) for a wide range of parameters.
It can describes both, the Kondo regime and the mixed-valence regime
(where the dot level is close to $E_F$, $-\Gamma_0\lesssim
\varepsilon_d\lesssim 0$).  However, in this approach the width of the
Kondo resonance decreases algebraically instead of having an
exponentional decay.  Then, we elaborate as well on a numerical
renormalization group analysis, which leads to \emph{nonperturbative}
results for all the regimes listed above.

\section{DOS splitting and TMR}
\begin{figure}\centering
\includegraphics*[width=85mm]{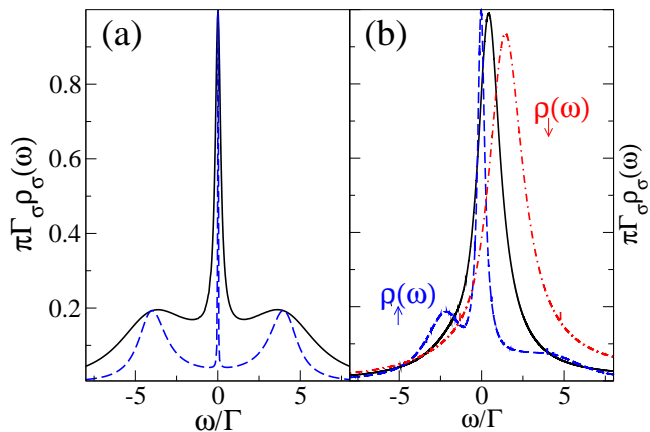}
\caption{(Color online). (a) Density of states for the symmetric
  Anderson model, $\varepsilon_d=-U/2$ at $T=0$ with
  $U/\pi\Gamma_0=2.5$. Solid line for $p=0$, and dashed line for $p=0.6$
  ($\Gamma_\uparrow\rho_\uparrow=\Gamma_\downarrow\rho_\downarrow$). (b)
  Density of states for the asymmetric Anderson model corresponding to
  the mixed valence regime, $\varepsilon_d=-\Gamma_0$ at $T=0$. The
  nonmagnetic case corresponds to the solid line. Spin up and down
  density of states for $p=0.6$ are shown with the dashed lines.}
\label{fig2}
\end{figure}

As indicated above, the Kondo resonance for a quantum dot coupled to two
ferromagnets with parallel magnetizations, splits away from the
symmetric case ($\varepsilon_d\neq U/2$) where charge fluctuations are
important. Figure~\ref{fig2} shows our results using the interpolative
$U$-finite perturbation theory including magnetic leads. The DOS for the
symmetric Anderson model is plotted in Fig.~\ref{fig2}(a) for the
unpolarized case $p=0$ and for nonzero polarization $p=0.6$ in the P
configuration. For unpolarized leads, the DOS shows the usual Kondo
resonance reaching the \emph{unitary limit} and two broad peaks at $\pm
U/2$ corresponding to the two mean-field (electron-like and hole-like)
peaks. The main effect of the polarized reservoirs is to make the Kondo
resonance narrower but keeping the same DOS height at $E_F$; i.e., for
$\varepsilon_d=-U/2$ the lead magnetizations preserve the unitary limit.

The physical scenario changes dramatically when charge fluctuations are
important as in the mixed valence regime. The solid line in
Fig.~\ref{fig2}(b) corresponds to $\varepsilon_d=-\Gamma_0$ for $p=0$.
The DOS displays a strongly renormalized level by the charge
fluctuations close to $E_F$ with no evidence of the mean-field peaks.
For a finite spin polarization $p=0.6$, Fig.~\ref{fig2}(b) depicts both
the spin up $\rho_{\uparrow}(\omega)$ and the spin down
$\rho_{\downarrow}(\omega)$ contributions to the local DOS
$\rho(\omega)=\rho_{\uparrow}(\omega)+\rho_{\downarrow}(\omega)$. Here,
$ \rho_{\downarrow \, (\uparrow)}(\omega)$, moves toward positive
(negative) frequencies. As a result, $\rho(\omega)$ shows a splitting at
low frequencies and the quantum occupations per spin change appreciably:
$\langle \hat{n}_{\uparrow}(p=0)\rangle<\langle
\hat{n}_{\uparrow}(p=0.6)\rangle$ and $\langle
\hat{n}_{\downarrow}(p=0)\rangle>\langle
\hat{n}_{\downarrow}(p=0.6)\rangle$. This demonstrates the sensitivity
of the spintronic Kondo effect to variations of the external gate
voltage.

\begin{figure}\centering
\includegraphics*[width=85mm]{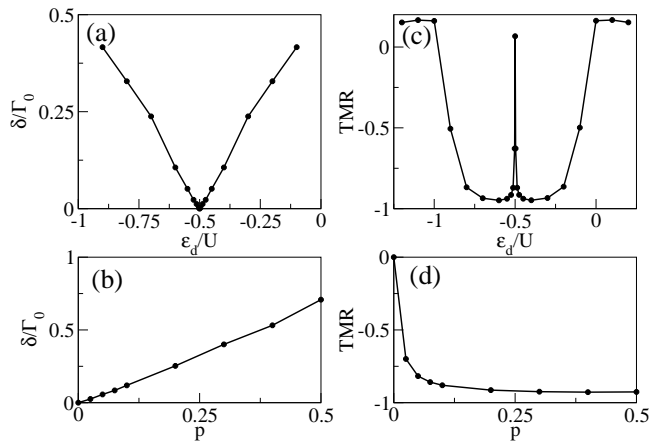}
\caption{(a) Splitting $\delta$ of the Kondo peak
  as a function of $\varepsilon_d$ for $p=0.25$ and $U=0.4D$ ($D$ is the
  continuum bandwidth).  (b) $\delta$ versus $p$ for
  $\varepsilon_d=-0.1D$ and $U=0.45D$. (c) TMR versus $p$ for
  $\varepsilon_d=-0.1D$ and $U=0.4D$. (d) TMR as a function of
  $\varepsilon_d$ for $p=0.25$ and $U=0.4D$. In all cases
  $\Gamma_0=0.02D$.}
\label{nrg}
\end{figure}

We now use a numerical renormalization group calculation to investigate
both the linear conductance and the splitting in the total DOS as a
function of the gate voltage $\varepsilon_{d}$ and the polarization of
the leads $p$. Figure~\ref{nrg}(a) shows the splitting $\delta$ of the
Kondo peak as a function of the gate voltage~\cite{choi04}. It increases
linearly from zero as the gate moves away from the symmetric point
$\varepsilon_d=-U/2$.  In terms of the lead polarization $p$ (the
arrangement is parallel), $\delta$ is linear as well [see
Fig.~\ref{nrg}(b)].

In Fig.~\ref{nrg}(d) we plot the TMR defined as
\begin{equation}
{\rm TMR}=\frac{G_{\rm P}-G_{\rm AP}}{G_{\rm AP}} \,,
\end{equation}
where $G_{\rm P}$ ($G_{\rm AP}$) is the linear conductance in the P (AP)
case. For the symmetric Anderson model, the Kondo effect survives even
for a finite value of polarization $|p|<1$. Then, at
$\varepsilon_d=-U/2$ we find that ${\rm TMR}= p^2/(1-p^2)$, in excellent
agreement with the numerical result. Away from the symmetric point,
i.e., $\varepsilon_d\neq-U/2$, $g^{P}$ gets strongly suppressed as $p$
increases. Then, the system exhibits a strong negative TMR [see
Fig.~\ref{nrg}(c)].  As a result, we predict a {\em sharp} peak of the
TMR by varying the gate potential [see Fig.~\ref{nrg}(d)].  The origin
of this peak is exclusively due to the particularities of the spintronic
Kondo effect.

We have so far discussed the case of a quantum dot symmetrically coupled
to the leads, i.e., $\Gamma_{L}=\Gamma_R=\Gamma_0$. As we have seen, the
P configuration leads to nontrivial effects in the transport properties
of the dot for $\varepsilon_{d}\neq -U/2$ since
$\varepsilon_{d,\uparrow}$ and $\varepsilon_{d,\downarrow}$ are not
equally coupled to the leads whereas for the AP configuration both
$\varepsilon_{d,\uparrow}$ and $\varepsilon_{d,\downarrow}$ are
renormalized by the Kondo correlations in the same manner. This scenario
is modified when we take an asymmetric quantum dot, $\Gamma_{L}\neq
\Gamma_{R}$. In this case, both configurations (P and AP) give rise to a
split DOS since $\varepsilon_{d\uparrow}$ is coupled to the leads with
$\Gamma_{L\uparrow}+\Gamma_{R{\uparrow}}$ unlike
$\varepsilon_{d\downarrow}$ (with
$\Gamma_{L\downarrow}+\Gamma_{R{\downarrow}}$). In general, there will
be splitting provided $\Gamma_{L\uparrow}+\Gamma_{R_{\uparrow}}\neq
\Gamma_{L\downarrow}+\Gamma_{R{\downarrow}}$

\section{Shot noise}
The shot noise are the dynamical fluctuations of the current
(current-current correlations) that appear in electric conductors due to
the quantization of the charge. Research on shot noise in mesoscopic
physics has developed into a fruitful area of research~\cite{bla00}.
Nevertheless, there have hitherto been very few attempts to investigate
shot noise in Kondo
impurities~\cite{mei02,don02,avi03,aon03,lop03,lop03b,cc}.

Here, we consider the the noise power (the Fourier transform of the time
correlator of the electric current) at zero frequency:
\begin{equation}
S_{\alpha\beta}(\omega=0)=2 \int d\tau \,
\langle \{\delta\hat{I}_\alpha(\tau),
\delta\hat{I}_\beta(0) \} \rangle
= 2 \int d\tau \, \left[
\langle \{\hat{I}_\alpha(\tau),
\hat{I}_\beta(0) \} \rangle -
\langle \hat{I}_\alpha \rangle\langle \hat{I}_\beta \rangle
\right] \,,
\label{noise1}
\end{equation}
where $\delta\hat{I}_\alpha=\hat{I}_\alpha-{I}_\alpha$ describes the
fluctuations of the current away from its average value
$I_\alpha=\langle \hat{I}_\alpha \rangle$.  We shall work at $T=0$ so
that the current will fluctuate due to quantum fluctuations only (we
disregard thermal fluctuations).

Within slave-boson mean-field theory, the shot noise in a two-terminal
geometry is shown~\cite{lop03b} to have the well known expression $S\sim
\tilde{\mathcal{T}}(1-\tilde{\mathcal{T}})$, i.e., the conventional
result for the partition noise but with {\em renormalized} transmissions
$\tilde{\mathcal{T}}$.  This is valid as long as we restrict ourselves
to the Fermi-liquid fixed point of the Kondo problem.
\begin{figure}\centering
\includegraphics*[width=85mm]{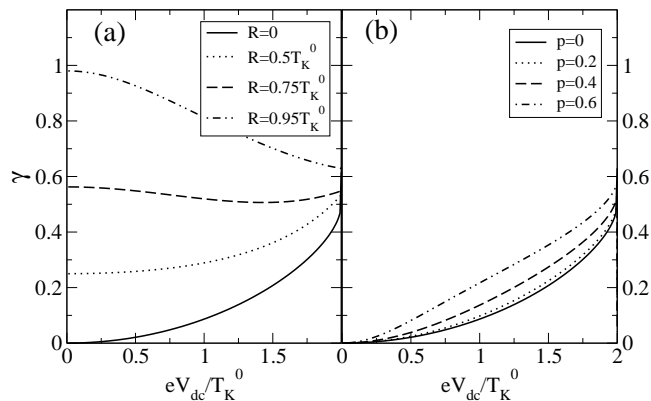}
\caption{Fano factor $\gamma$ versus applied dc voltage $V_{\rm dc}$
  of a quantum dot in the Kondo regime attached to ferromagnetic leads
  and in the presence of spin flip scattering with rate $R$. Voltage is
  units of the Kondo temperature $k_B
  T_K^0=D\exp{(-\pi|\varepsilon_d|/2\Gamma)}$.  We take
  $\varepsilon_d=-6\Gamma$ and $D=100\Gamma$.  (a) Lead polarizations
  are $p_L=p_R=0.5$. (b) $R=0$ and $p=p_L=p_R$ (parallel alignment).}
\label{noise}
\end{figure} 

It is customary to define the Fano factor:
\begin{equation}
\gamma=\frac{S(0)}{2e\langle I\rangle}\,.
\end{equation}
Since we are dealing with a two-terminal system, we have dropped the
lead indices.  Now, for a classical conductor with no correlations, the
Fano factor equals 1 (Poissonian limit). Deviations of this limit are
usually due to the application of Pauli principle or to the effect of
strong electron-electron interactions as those giving rise to the Kondo
effect.  In Fig.~\ref{noise}(a), we plot the influence of spin flips in
$\gamma$. The polarization in the leads is taken as $p_L=p_R=0.5$. At
low bias, $\gamma$ behaves as $1-\tilde{\mathcal{T}}(E_F)$~\cite{bla00}.
Since for $R=0$ the Kondo resonance achieves the unitary limit at zero
bias, the Fano factor is completely suppressed down to zero.  As $R$
increases, spin flips induce decoherence in the correlated motion of the
electrons which leads to the singlet formation. Hence, the transmission
at $E_F$ departs from its unitary limit and, as a consequence, $\gamma$
increases at zero bias ($V_{\rm dc}=0$). For larger voltages, notice
that at $R=0$ we recover the limit $\gamma=1/2$ of a double-barrier
resonant system~\cite{bla00}.  For $R>0$ the behavior of $\gamma$ at
larger $V_{\rm dc}$ depends on the particular two-peak structure of the
transmission~\cite{lop03}.

In Fig.~\ref{noise}(b), we calculate the Fano factor for $R=0$ and
different values of the lead magnetizations (P configuration,
$p=p_L=p_R$).  We find that $\gamma$ increases with the polarization $p$
at a given voltage bias (except at $V_{\rm dc}=0$).  This is caused by
the suppression of the Kondo effect for $V_{\rm dc}>0$. For AP
alignments (not shown here), $\gamma$ increases less rapidly due to the
independence of $T_K$ on the lead polarization.

Notice that in slave-boson mean-field theories the fluctuations of the
boson field are neglected. However, we do not expect large deviations
from the results reported here when $T\ll T_K$.  The boson fluctuations
will evidently become important as temperature approaches $T_K$.

\section{Possible future advances}
We have demonstrated that rich physics appears when the formation of the
Kondo state in a quantum dot competes with the presence of
spin-polarized tunneling currents and spin-flip processes.  We discuss
now possible extensions of the theory that could dramatically alter the
effects exposed above. We are confident that the field will still offer
unexpected results and that, consequently, future calculations and
experiments will be full of rewards.

\subsection{Magnon-assisted transport}
We have thus far considered metallic free-electron ferromagnets as the
injecting and receiving contacts. In reality, transition-metal
electrodes are described by exchange Hamiltonians. In these models, it
is assumed that electrical conduction is carried by itinerant
$s$-electrons while (insulator) magnetism is caused by a different
group: localized $d$-electrons. Interaction between free electrons and
localized moments gives rise to \emph{electron-magnon}
coupling~\cite{her73}.

It has been suggested that magnon-assisted tunneling in magnetic tunnel
junctions may lower the TMR as a function of the bias
voltage~\cite{zha97,bra98}, giving rise to a zero-bias anomaly in
nonlinear current--voltage characteristics~\cite{moo95} (see, e.g.,
Ref.~\cite{Tsymbal03a} for a more detailed review on the subject) .
The peak width is given by the energy involved in the spin excitations,
which is of the order of the Curie temperature ($T_C$) of the metal. As
$T_C\gg T_K$, one would naively expect that the Kondo effect will be
always destroyed by emission and absorption of magnons via spin-flip
processes. Within the tunneling Hamiltonian formalism, we replace ${\cal
  H}_{\rm leads}$ in Eq.~(\ref{eq_h}) with
\begin{equation}
{\cal H}_{\rm leads}=\sum_{k\alpha\sigma}
\varepsilon_{k\alpha\sigma}c_{k\alpha\sigma}^\dagger 
c_{k\alpha\sigma} - J_{dd} \sum_{\langle i,j\rangle \alpha}
\vec{S}_{\alpha i}
\cdot \vec{S}_{\alpha j} - J_{sd} \sum_{i} \psi_{\alpha\sigma}^\dagger
(\vec{s}_{\sigma \sigma'}\cdot\vec{S}_{\alpha i}) \psi_{\alpha\sigma'} \,,
\end{equation}
where the first term describes the conduction band electrons, the second
term is the Heisenberg interaction between localized moments
$\vec{S}_{\alpha i},\vec{S}_{\alpha j}$ at neighboring sites $i$ and $j$
of lead $\alpha$ and the third term is the interaction between a
localized moment at site $i$ and an itinerant electron with creation
operator $\psi_{\alpha\sigma}^\dagger=\sum_{\vec{k}} e^{-i\vec{k_\alpha}
  \cdot \vec{r}_{\alpha i}} c_{k\alpha\sigma}^\dagger$. After applying
the Holstein-Primakoff transformation, the ferromagnet low-energy spin
excitations can be written in terms of a bosonic collective bath
(magnons), each carrying a magnetic moment $e\hbar/mc$. Thus,
electron-magnon interaction ${\cal H}_{em}$ at the interface induces
spin mixing~\cite{zha97}:
\begin{equation}\label{eq_hem}
{\cal H}_{em}=-\tilde{J}_{sd} \sum_{k,k',q} [
c_{k\alpha\uparrow}^\dagger c_{k\alpha\downarrow}
(a_{q}^\dagger +a_q) +\rm{H.c.}]\,,
\end{equation}
where the $\tilde{J}_{sd}$ is a renormalized coupling constant (which is
taken as momentum independent for simplicity) and $a_q^\dagger\sim
N^{-1/2} \sum_{\alpha i} \exp(-i \vec{q}\cdot \vec{r}_{\alpha i})
S_i^{-}$ is the magnon creation operator.

In Eq.~(\ref{eq_hem}), we have written down only the spin-flip part of
the electron-magnon interaction. It will lead to nontrivial correlations
when combined with the Hamiltonian ${\cal H}_{\rm dot}$ of a
Coulomb-blockade quantum dot in Eq.~(\ref{eq_h}).  In fact, ${\cal
  H}_{em}$ involves spin-flip inelastic transitions (the term
proportional to $R$ in Eq.~(\ref{eq_hr}) is simply elastic), inducing
decoherence in the Kondo resonance by means of emission (dominant at low
$T$) and absorption of magnons of energy $\hbar \omega_q$.  (At low
temperatures, $\omega_q$ depends approximately on $q$ in a quadratic
way, $\omega_q\sim q^2$).  Furthermore, subtle out of equilibrium
effects such as interlayer exchange interaction~\cite{hei99} may arise
as well.

We should mention that magnon excitations may also act as a {\em
  dissipative} bath. Since the standard spectral density for a
three-dimensional (cubic) ferromagnet goes as
$\omega^{1/2}$~\cite{bra98}, we would deal with a {\em subohmic} bath.
(The spectrum of the bath is cut off by a maximum magnon frequency due,
e.g., to an anisotropy energy).  For comparison, the DOS of magnons in
antiferromagnetic two-dimensional systems varies as $\omega$, which
amounts to an {\em ohmic} bath.

\subsection{Double quantum dots}
Double-quantum-dot (DQD) systems have recently attracted much attention
since that they form the simplest artificial systems showing
molecule-like correlations at the nanoscale. As a consequence, a DQD has
been proposed as a basic constituent of a solid-state quantum
computer~\cite{book}.

\begin{figure}\centering
\includegraphics*[width=85mm]{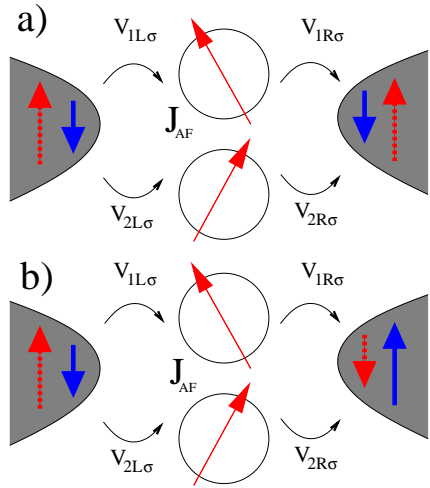}
\caption{(Color online). Schematic picture of the parallel double
  quantum dot attached to ferromagentic leads. $J_{\rm AF}$ is the
  interdot exchange coupling between the spin of the dots. (a) Parallel
  configuration of the leads polarization. Dashed arrows correspond to
  the majority spins and solid arrows are for minority ones. (b)
  Antiparallel alignment of the leads polarization.}
\label{dqd}
\end{figure} 

Two quantum dots can be coupled either in series or in parallel,
allowing for tunneling and capacitive couplings in between. As far as
Kondo physics is concerned, a DQD may be regarded as an artificial
realization of the {\em two-impurity Kondo
  problem}~\cite{jay,geo,ros02,ros04,craig}.  It consists of two Kondo
impurities with spin $\vec{S}_1$ and $\vec{S}_2$ interacting via an
antiferromagnetic (AF) exchange coupling, $J_{\rm AF}\vec{S}_1 \cdot
\vec{S}_2$:
\begin{equation}
{\cal H}_{\rm AF}=J_{\rm AF} \sum_{\sigma \sigma'}
d_{1\sigma}^\dagger d_{1\sigma'}
d_{2\sigma'}^\dagger d_{2\sigma} \,,
\end{equation}
where $d_{1\sigma}^\dagger (d_{2\sigma'}^\dagger)$ creates an electron
at dot 1 (2) with spin $\sigma$ ($\sigma'$).  It has been shown that the
ratio $J_{\rm AF}/ T_K$ determines the ground state of the system. In
particular, when $J_{\rm AF}\gg T_K$ the two dots are locked into a
antiferromagnetic singlet state whereas for $J_{\rm AF}\ll T_K$ each dot
forms its own Kondo state with continuum electrons in the leads.  The
critical value at which the transition from the Kondo state (KS) to the
AF phase takes place can be obtained by comparing their ground state
energies. Thus, the critical point depends on the Kondo temperature for
each dot ($T_K^{1},T^2_{K}$) as follows
\begin{equation}
\left(\frac{J_{\rm AF}}{T_K^{1}}\right)_c
=\frac{4}{\pi}\left(1+\frac{T^2_{K}}{T_{K}^1}\right)\,. 
\end{equation}
For a symmetrically coupled DQD with a common gate
$\varepsilon_1=\varepsilon_2$ one has $(J_{\rm AF}/T_K)=8/\pi$. In
general for $0\leq T^2_{K}\leq T_{K}^1$ we have $4/\pi \leq
(I/T_K^{1})_c\leq 8/\pi$. Since $T^1_{K}$, and $T^2_{K}$ depend
exponentially on the tunneling couplings and the level positions, a
small asymmetry between these parameters induces a huge change in the
ratio $T^2_{K}/T_{K}^1$.

Let us consider a parallel DQD (see Fig.~\ref{dqd}) connected to two
ferromagnetic leads. The Kondo temperature for the dot $i\in\{1,2\}$
depends on the configuration of the polarization of the leads (parallel
or antiparallel). To simplify, we take the same polarization for the
leads $p>0$ (P alignment) and identical dots
$T^2_{K}(p)=T_{K}^1(p)=T_K(p)$. Now the transition from
KS$\rightarrow$AF singlet state is achieved more easily by increasing
$p$. We keep the antiferromagnetic coupling fixed $J_{\rm AF}\ll
T_K(p=0)$ and vary $p$. In this way, $T_K(p)$ becomes smaller leading to
a weaker Kondo effect. Here, the width of the zero-bias anomaly (ZBA)
decreases with $p$ and the conductance reaches the unitary limit as in
the case of a single dot~\cite{lop03,choi04}.  By further increasing
$p$, the AF coupling is much stronger than the Kondo scale leading to
the transition ${\rm Kondo}\rightarrow {\rm AF}$ when $J_{\rm
  AF}/T_K(p)>8/\pi$.

\subsection{Higher symmetry Kondo states}
In DQD systems with a strong interdot Coulomb interaction, the total
charge allowed in both dots at the same time is just one electron. As a
consequence, there are four ground states with the same energy, namely,
$\{1\uparrow, 1\downarrow, 2\uparrow, 2\downarrow \}$.  Quantum
fluctuations between these states due to coupling to the leads yield, in
the low temperature limit, a highly correlated state with SU(4)
symmetry~\cite{bor03}.  Note that these fluctuations do not involve only
spin flips in the DQD (spin Kondo effect) but also flips in the {\em
  orbital} sector~\cite{zar03}. To describe these new processes, we
define the {\em pseudospin} as a fictitious spin that points along
$+$($-$)$z$ when the electron lies at the dot 1(2). Then, it is shown
that the spin Kondo state becomes intermingled with a pseudospin Kondo
state, giving rise to a complete entanglement between the spin and
charge degrees of freedom.

The spin-pseudospin entanglement develops from the correlated tunneling
that involves a flip of the spin {\em and} the pseudospin of the DQD
system at the same time. Technically, it arises from the
Schrieffer-Wolff transformation~\cite{hew93} which maps the DQD
Hamiltonian into an effective exchange coupling between the localized
spin and pseudospin and the conduction electrons. In the DQD
Hamiltonian, one replaces ${\cal H}_{\rm dot}$ in Eq.~(\ref{eq_h}) with
the Hamiltonian of two dots plus a charging energy $U_{12}$ between
them:
\begin{equation}
{\cal H}_{\rm inter}=\sum_{i=1,2} U_{12}\hat{n}_{1}\hat{n}_{2}\,.
\end{equation}
$\hat{n}_i$ denotes the occupation number on dot $i$.  We denote with
$\vec{T}$ the pseudospin operator in the DQD system. The resulting Kondo
Hamiltonian formally reads:
\begin{equation}\label{eq_su4}
{\cal H}_K^{\rm SU(4)}=J^{\rm SU(4)}
\vec{S}\cdot(\psi^\dag\vec{\sigma}\vec{\tau}\psi)\cdot\vec{T}\,,
\end{equation}
where $\psi^\dag=\sum_{k}\psi_{k}$ is the field operator with
$\psi_{k}=[c^\dagger_{e,k,\uparrow}, c^\dagger_{o,k,\uparrow},
c^\dagger_{e,k,\downarrow}, c^\dagger_{o,k,\downarrow}]$ a spinor in the
representation of even and odd channels of the lead
operators~\cite{su4}. In Eq.~(\ref{eq_su4}) $J^{\rm SU(4)}$ is a
coupling constant which goes to the strongly fixed point in the flow
diagram.  We recall that the SU(4) Kondo state takes place provided
there are two conduction channels (described by the matrix
$\vec{\tau}$). The latter equation explains the entanglement between the
spin and the orbital electronic degrees of freedom.

The transport properties of a SU(4) Kondo state strongly differ from the
conventional SU(2) Kondo state, both at equilibrium and out of
equilibrium~\cite{su4}.  First, the Kondo temperature inferred from
Eq.~(\ref{eq_su4}) is largely enhanced (around 200 times) as compared to
$T_K$ of a spin Kondo system~\cite{boe02}.  This means that the
differential conductance peak (which mimics the DQD density of states)
becomes greatly broadened in a transition from the SU(2) to the SU(4)
Kondo physics. Such a transition can be tuned with a magnetic flux in an
Aharonov-Bohm interferometer with one dot at each arm~\cite{su4}.
Second, the Kondo resonance is no longer peaked at $E_F$ but at $\sim
E_F+T_K$ to fulfill the Friedel-Langreth sum rule~\cite{hew93}.

How would spintronic transport modify a SU(4) Kondo resonance?  In the
presence of ferromagnetic leads and away from the particle-hole symmetry
point, we expect the spin part of the Kondo state to slowly vanish with
increasing lead polarization. Experimentally, one would see a splitting
of the Kondo resonance into {\em three} peaks.  The centered peak would
still correspond to the pseudospin Kondo state, which is not sensitive
to the magnetization at the leads. At the same time, $T_K^{\rm SU(4)}$
decreases but the linear conductance would increase since the SU(2)
Kondo resonance associated to the orbital Kondo effect peaks at $E_F$
again. To further destroy the pseudospin Kondo state, two possibilities
emerge from an analogy with the spin case. First, one allows for
tunneling between the dots, which breaks the fourfold degeneracy
favoring the formation of a bonding (symmetric) state between the dots.
Then, interdot tunneling acts as an external Zeeman splitting in the
spin sector. Second, one could consider asymmetric couplings of the DQD
to the leads; e.g., the DQD system may be coupled strongly to the left
lead, $\Gamma_L>\Gamma_R$.  This way, we regard the leads as {\em
  pseudospin polarized} much like the spin-dependent tunneling due to
ferromagnetic leads in the spin case. We may even define the pseudospin
polarization for each spin species as
\begin{equation}\label{eq_pp}
p_\sigma=
\frac{\Gamma_{\sigma L}-\Gamma_{\sigma R}}{
\Gamma_{\sigma L}+\Gamma_{\sigma R}}
\,,
\end{equation}
[cf. Eq.~(\ref{eq_p})]. At this point, more calculations are needed to
further exploit this analogy between spins and pseudospins, which may
give rise to a unifying picture of the influence of real and pseudo-spin
polarized leads in the transport through quantum-dot structures.

\section{Conclusions and experimental relevance}
We have investigated the spintronic properties of a quantum dot in the
Kondo regime. We have considered a dot attached to ferromagnetic leads
and in the presence of intradot spin flip scattering.  Using both
perturbation theory in the on-site interaction and the NRG method, we
have shown that the Kondo effect is not necessarily suppressed by the
spin polarizations of the leads: for the symmetric Anderson model, where
charge fluctuations are completely suppressed, the Kondo effect is
robust even for finite polarizations.  For the asymmetric Anderson
model, the Kondo peak does split into two.  This is due to the presence
or absence of particle-hole symmetry.  In the presence of particle-hole
symmetry, the Kondo peak at the Fermi level remains unsplit even at
finite polarizations and the linear conductance achieves the unitary
limit. This remains true as long as only spin fluctuations are present
in the QD.  On the contrary, when particle-hole symmetry is absent, the
conductance is suppressed due to the visible splitting of the Kondo
peak. Since the Kondo resonance is mostly unaltered for antiparallel
magnetizations, we have calculated the TMR in the Kondo, mixed-valence,
and empty-level regimes. The TMR shows a characteristic behavior for
each of them. In addition, we have shown that the TMR is strongly
affected in the presence of spin flip processes.

We have studied the form of the shot noise when charge fluctuations are
completely suppressed. We have shown that the Fano factor approaches the
Poissonian limit when the spin flip scattering rate is of the order of
the Kondo temperature. For parallel arrangements, the Fano factor
enhances with increasing lead polarizations.

Moreover, we have suggested and discussed possible new advances in this
field such as the influence of magnons in the Kondo state of a quantum
dot, the effect of spin-polarized currents on double-quantum-dot systems
mimicking the two-impurity problem and on more exotic Kondo states with
higher symmetry.

The physics addressed in this paper is realistic and can be visible
within the scope of present techniques as the energies we treat are
within the Kondo scale. In particular, a change has been detected in the
resistivity of a Kondo alloy due to spin-polarized
currents~\cite{Taniyama03a}.  Furthermore, it is already possible to
attach ferromagnetic leads to a carbon nanotube~\cite{Tsukagoshi99a},
and a carbon-nanotube quantum dot has been shown to display Kondo
physics below an unusually high temperature~\cite{Nygard00a}.
Finally, a quantum dot coupled to ferromagnetic electrodes has been
proposed as a promising candidate for spin injection devices, and
studied experimentally both in the Coulomb blockade
regime~\cite{Deshmukh02a} and in the Kondo
regime\cite{Pasupathy04a}.

\section{Acknowledgements}
We thank R.~Aguado, M.~B\"uttiker, T.~Costi, L.~Glazman, K.~Kang,
K.~Le~Hur, G.~Platero, P.~Simon, and E.V.~Sukhorukov for valuable
comments and discussions on this subject.  D.S. and R.L. were supported
by the the EU RTN under Contract No. HPRN-CT-2000-00144, Nanoscale
Dynamics, and by the Spanish MECD.  M.-S.C.\ acknowledges supports from
the SKORE-A program, and from the eSSC at POSTECH.

\end{document}